\documentclass[a4paper,11pt]{article}
\usepackage{pos}
\usepackage{orcidlink}
\usepackage{microtype}
\usepackage{placeins}
\hypersetup{
    pdftitle={A new view of the cosmic-ray knee after LHAASO},
    pdfauthor={Carmelo Evoli},
    colorlinks=true,
    citecolor=magenta,
    linkcolor=magenta,
    pdfpagemode=UseNone
}
\graphicspath{{figures/}}

\title{A new view of the cosmic-ray knee after LHAASO}
\ShortTitle{The cosmic-ray knee after LHAASO}
\author*[a,b]{Carmelo Evoli\,\orcidlink{0000-0002-6023-5253}}
\renewcommand{\printHeadAuthors}{Carmelo Evoli}
\affiliation[a]{Gran Sasso Science Institute (GSSI), Viale Francesco Crispi 7, 67100 L’Aquila, Italy}
\affiliation[b]{INFN-Laboratori Nazionali del Gran Sasso (LNGS), via G. Acitelli 22, 67100 Assergi (AQ), Italy}
\emailAdd{carmelo.evoli@gssi.it}

\abstract{Recent LHAASO measurements offer a new view of the cosmic-ray
knee by combining the all-particle spectrum and mean logarithmic mass
with proton and light-component spectra across the PeV region.
This contribution reviews these results and their implications for
Galactic acceleration and transport. Approximately charge-ordered
proton and helium knees coexist with different spectral evolution,
while their turnover shapes constrain source populations and escape
mechanisms. These measurements anchor an elemental ``ladder'' that,
extended to heavier nuclei, will be crucial for establishing how far
the Galactic component persists and where the Galactic--extragalactic
transition takes place.}

\FullConference{International Symposium on Very High Energy Cosmic Ray Interactions (ISVHECRI2026)\\
5-10 July, 2026\\
Lecce, Italy\\}

\newcommand{\PeV}{\,\mathrm{PeV}}
\newcommand{\PV}{\,\mathrm{PV}}
\newcommand{\rig}{\mathcal R}

\begin{document}
\maketitle


\section{Introduction}
\label{sec:intro}

The ``knee''---a steepening of the all-particle spectrum at a few
PeV---remains a central puzzle in cosmic-ray physics.
Often associated with the high-energy end of
the Galactic spectrum, it provides a key landmark for the
Galactic--extragalactic transition~\cite{Bluemer2009}. In a rigidity-ordered
picture, a Galactic population fades through successive elemental
knees, with heavier nuclei reaching higher total energies. Tracking
composition helps establish how far the Galactic
contribution extends and whether additional Galactic sources are needed
before an extragalactic component becomes dominant.

The origin of this feature is usually explored within two broad
frameworks. In acceleration models, the finite lifetime, size, and
magnetic field of the sources set the maximum energy of accelerated particles.
In transport models, the knee appears when the energy dependence of
Galactic escape changes, so that the residence time decreases more
rapidly above a characteristic rigidity. Both mechanisms can produce a
steepening and approximately charge-ordered elemental knees, leaving
some degeneracy even when composition measurements are available.
Nevertheless, as discussed below, the detailed shape of an elemental
spectrum can provide strong constraints on the proposed mechanisms.

LHAASO has brought a qualitatively new perspective to these questions.
Its precise measurements of the all-particle spectrum and mean
logarithmic mass, $\langle\ln A\rangle$, trace both the spectral
steepening and the accompanying evolution of composition across the
knee~\cite{LHAASOall}. Complementary measurements of the proton spectrum
and the light component, consisting of protons and helium, connect the
behavior of the total flux to that of its principal light
constituents~\cite{LHAASOp,LHAASOHe}.

Together, these observations reveal spectral hardenings before the
elemental knees, changes in the relative proton and helium abundances,
and a turn toward a heavier mean composition above the all-particle
knee. In my review talk\footnote{\url{https://agenda.infn.it/event/47390/contributions/296052/}},
I discussed how these new LHAASO measurements
reshape our view of Galactic models of escape and source maximum
energies~\cite{Kachelriess2019}.

\section{LHAASO measurements: the all-particle knee and mean mass}
\label{sec:all}

I recap here the LHAASO measurements, starting with the joint measurement
of the all-particle spectrum and mean logarithmic mass by LHAASO-KM2A
over $0.3$--$30\PeV$
\cite{LHAASOall}. At the altitude of LHAASO, showers in the knee region
are observed close to their maximum development. KM2A measures their
electromagnetic and muonic signals, combining them in the energy estimator
$N_{e\mu}=N_e+2.8N_\mu$, with both signals evaluated between 40 and
$200\,\mathrm{m}$ from the shower axis. This combination reduces the
dependence of the reconstructed energy on the primary mass. At a given
energy, heavier nuclei produce more muons; simulations then relate
$\langle\ln N_\mu\rangle$ to the mean logarithmic mass,
$\langle\ln A\rangle=\sum_i f_i\ln A_i$, where $f_i$ is the fraction
of species $i$ in the total flux.

\begin{figure}[t]
 \centering
 \includegraphics[width=0.70\textwidth]{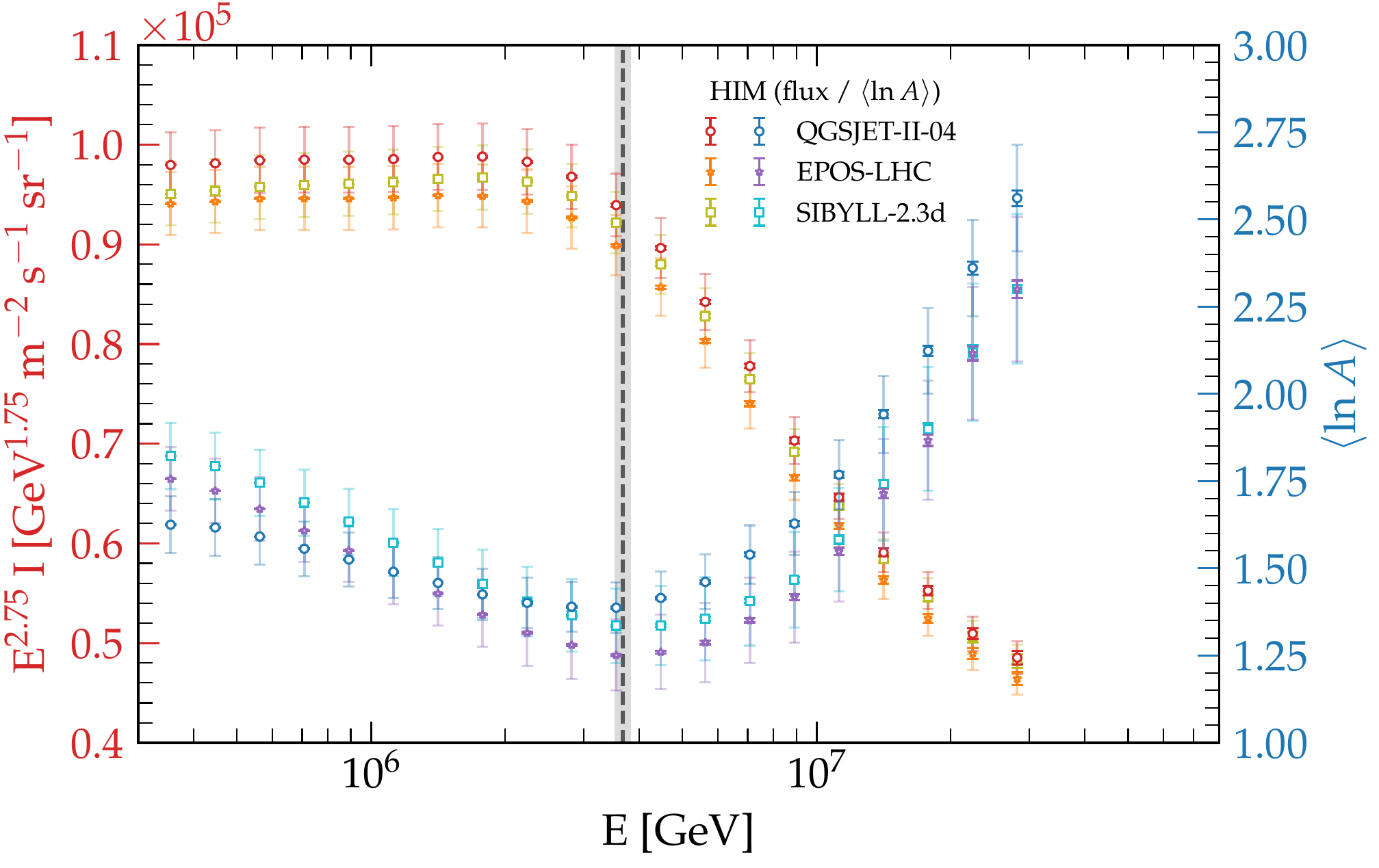}
 \caption{LHAASO all-particle intensity multiplied by $E^{2.75}$ (left
 axis) and mean logarithmic mass (right axis), for three hadronic models.
 The dashed line marks the knee at $3.67\PeV$ in
 Eq.~\ref{eq:allknee}, obtained with an energy calibration averaged over
 the three HIMs. The gray band shows its uncertainty, combining the quoted
 statistical and systematic errors in quadrature.
 Data from Ref.~\cite{LHAASOall}.}
 \label{fig:all}
\end{figure}

The all-particle spectrum, reconstructed using an energy calibration
averaged over the three hadronic interaction models (HIMs), is well described
by a smoothly broken power law, with a knee at
\begin{equation}
 E_{k,\rm all}=(3.67\pm0.05_{\rm stat}\pm0.15_{\rm sys})\PeV.
 \label{eq:allknee}
\end{equation}
Using the positive spectral index convention $J\propto E^{-\gamma}$,
the index changes from approximately $2.741$ to $3.128$, corresponding
to $\Delta\gamma=0.387\pm0.005_{\rm stat}\pm0.027_{\rm sys}$
\cite{LHAASOall}. Figure~\ref{fig:all} shows this steepening together
with the evolution of the mean mass.

The key result emerges when the two observables are read together.
The mean logarithmic mass decreases from about 1.7 at $0.3\PeV$ to
about 1.3 near $3\PeV$, indicating an increasing light-component
fraction, before turning toward a heavier composition. This minimum
lies near the all-particle knee, while the break of the fitted
function lies somewhat higher, at
$(5.36\pm0.08_{\rm stat}\pm0.46_{\rm sys})\PeV$~\cite{LHAASOall}.
Together with the elemental spectra below, these results favor a
proton-driven knee, contrasting with the helium-dominated expectation
extrapolated from CALET~\cite{CALET2023} and DAMPE~\cite{DAMPE2026}.

A remarkable advance of this measurement is its weak dependence on the
HIM, traditionally a major source of
uncertainty in air-shower studies of the knee. EPOS-LHC, QGSJET-II-04,
and SIBYLL 2.3d yield similar spectral and composition trends, with an
estimated HIM contribution of only about $\pm2.5\%$ to the all-particle
flux uncertainty and less than $\pm6\%$ to that on
$\langle\ln A\rangle$~\cite{LHAASOall}. The combined systematic
uncertainties, including the assessed HIM dependence, are about $4.5\%$
on the flux and $9\%$ on $\langle\ln A\rangle$. This stability gives a
firmer experimental basis for interpreting the spectral steepening
together with the reversal in mean mass.

\begin{figure}[t]
 \centering
 \includegraphics[width=0.70\textwidth]{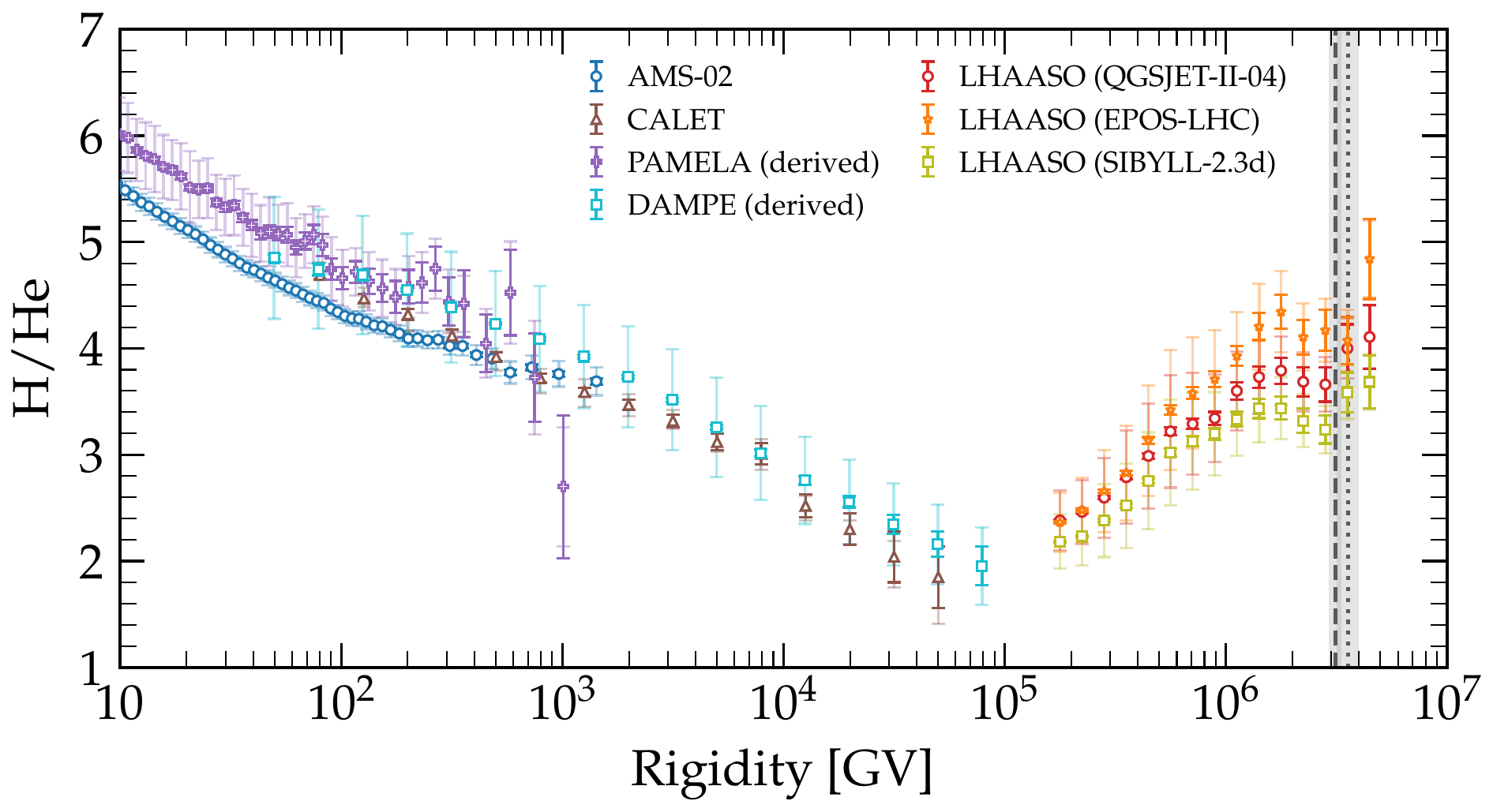}
 \caption{Proton-to-helium intensity ratio at equal rigidity from direct
 measurements~\cite{AMS2021,CALET2023,PAMELA2011,DAMPE2026} and
 LHAASO~\cite{LHAASOHe}. Opaque/pale bars show statistical/systematic
 errors. Dashed/dotted lines and bands mark the proton/helium knee
 rigidities from the joint fit of Section~\ref{sec:shape}.}
 \label{fig:ratio}
\end{figure}

\section{Resolving the proton and helium spectra}
\label{sec:elements}

The elemental spectra reveal how protons and helium contribute to the
all-particle knee and how their relative abundances evolve across it.
The hybrid KM2A--WFCTA analysis selects protons with approximately
90\% purity and achieves about 10\% energy resolution above $1\PeV$
\cite{LHAASOp}. The helium spectrum is derived by subtracting the
proton spectrum from the light-component spectrum,
$J_{\rm He}=J_{p+\mathrm{He}}-J_p$, on a common energy scale.
The helium analysis also updates the proton spectrum through iterative
contamination corrections~\cite{LHAASOHe}. The resulting spectra share
uncertainties that must be accounted for in fits and ratios.

\subsection{Protons: the dominant contribution to the knee}

The proton spectrum extends from about $0.15$ to $12\PeV$. It departs
from a simple extrapolation of the lower-energy spectrum, as it hardens, and
then steepens near $3.3\PeV$ \cite{LHAASOp}. In the published EPOS-LHC
fit, the hardening scale is about $0.34\PeV$, and the index changes
from about 2.71 to 2.51 before reaching about 3.5 above the knee.
The knee steepening is thus of order unity, substantially larger than
that of the all-particle spectrum.

The proximity of the proton and all-particle knees identifies protons
as the main driver: with about a third of the flux at the knee
(Section~\ref{sec:shape}), a proton steepening of $0.85$--$1.0$
can account for roughly $\Delta\gamma\simeq0.3$ of the measured $0.39$.
The preceding proton hardening is equally important: it increases the proton
contribution on the approach to the knee, reversing the trend inferred
from lower-energy direct measurements~\cite{CALET2023,DAMPE2026}. This motivates an additional
contribution from sources with acceleration properties different from
those of the population dominating at lower energies.

\subsection{Helium: a later knee and a different pre-knee evolution}

The helium spectrum covers $0.16$--$13\PeV$. It also exhibits hardening
centered near $1.1\PeV$, followed by a knee around $7\PeV$ and a
steepening of approximately one unit in spectral index
\cite{LHAASOHe}. Its lower-energy behavior differs from that of
protons: between about 0.3 and $1\PeV$, helium is softer. In total
energy, the measured intensities cross near $0.7\PeV$, where protons
overtake helium, and again near $5\PeV$, where helium overtakes protons.

The helium knee occurs at roughly twice the proton knee energy,
consistent with charge ordering, $E_{k,i}\simeq Z_i e\rig_k$, at a
common rigidity of order $3$--$4\PV$. This correspondence concerns
the positions of the knees; the different pre-knee evolutions already
show that it does not imply identical elemental spectra at equal
rigidity. Indeed, the proton-to-helium ratio at equal rigidity,
$J_p(e\rig)/[2J_{\rm He}(2e\rig)]$ in terms of the total-energy
intensities, reverses the decreasing trend inferred from direct measurements, rising toward
the knee (Fig.~\ref{fig:ratio}).

\subsection{Spectral shapes and composition at the knee}
\label{sec:shape}

Alongside the knee energy, the way the spectrum steepens provides
relevant physical information.
A broken power law approaches a new, finite spectral index above the
break, whereas an exponential cutoff produces continuing curvature.
To examine this distinction, I compare the two forms
\begin{equation}
 J_{\rm cut}(E)\propto E^{-\gamma_c}\exp(-E/E_{\rm cut}),\qquad
 J_{\rm br}(E)\propto E^{-\gamma_b}
 \left[1+\left(\frac{E}{E_b}\right)^{1/w}\right]^{-w\Delta\gamma}.
 \label{eq:knee-shapes}
\end{equation}
The first has a local index $\gamma_c+E/E_{\rm cut}$; the second
changes from $\gamma_b$ to $\gamma_b+\Delta\gamma$, with transition
width controlled by $w$. The cutoff scale $E_{\rm cut}$ and the break
energy $E_b$ consequently have different meanings.

\begin{figure}[t]
 \centering
 \includegraphics[width=0.49\textwidth]{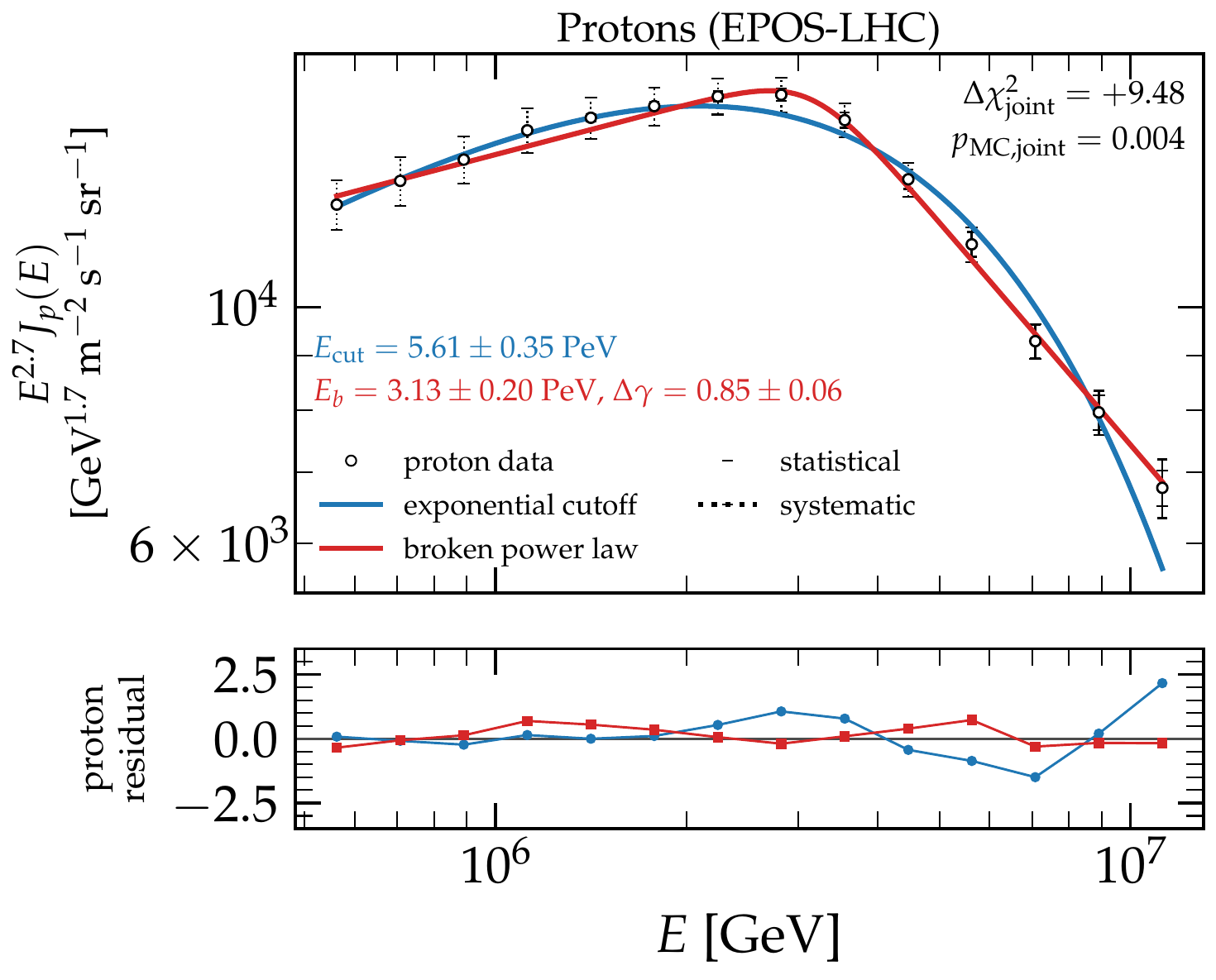}\hfill
 \includegraphics[width=0.49\textwidth]{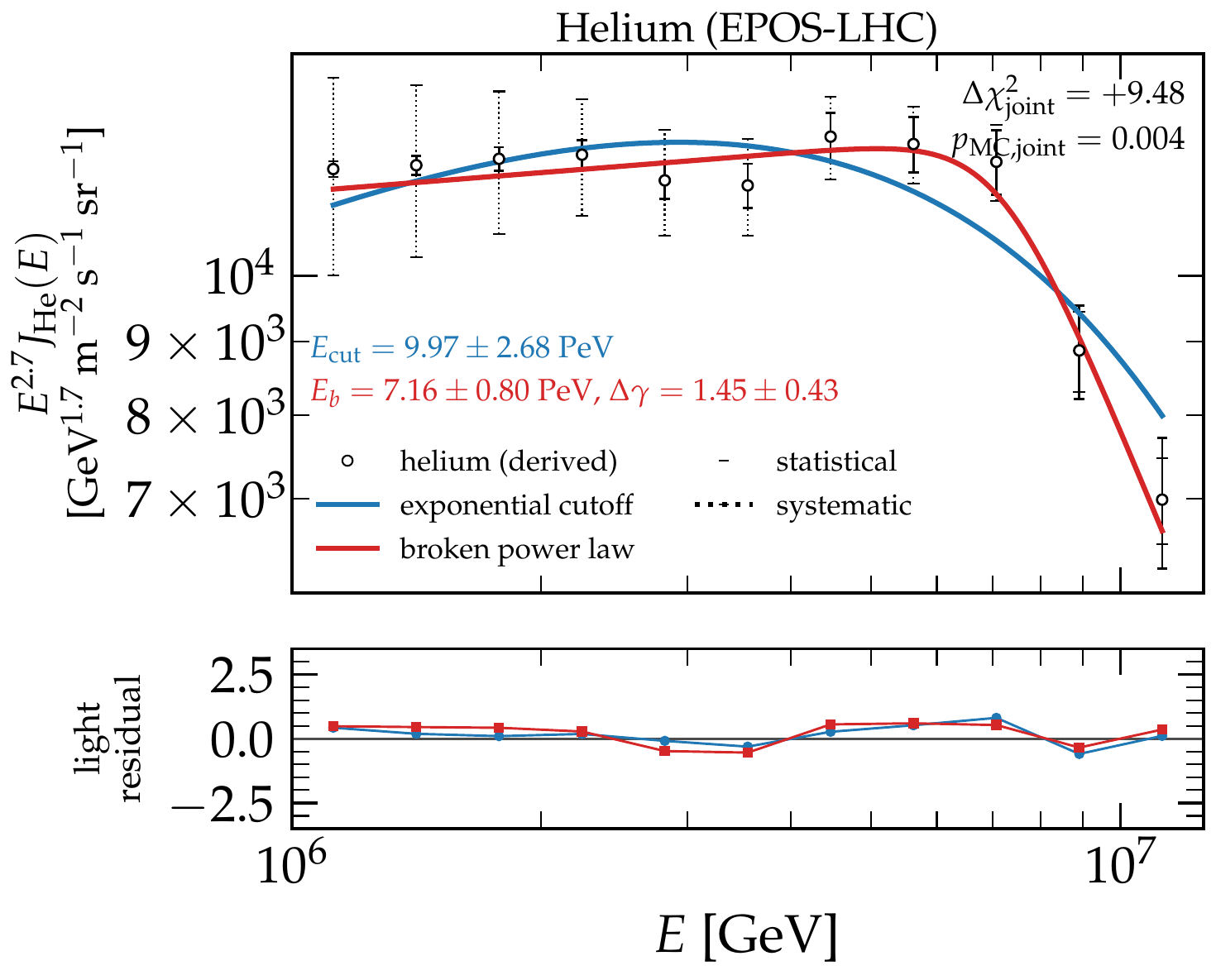}
 \caption{Joint fit to the EPOS-LHC proton and light-component spectra
 of Ref.~\cite{LHAASOHe}: exponential cutoffs (blue) and smoothly
 broken power laws with $w=0.1$ (red). Helium points (right) are
 derived and shown only for comparison. Solid/dotted bars:
 statistical/systematic errors. Lower panels: normalized residuals of
 the fitted proton (left) and light-component (right) data.}
 \label{fig:kneeshape}
\end{figure}

I jointly fit the EPOS-LHC proton and light-component
spectra~\cite{LHAASOHe} above $0.5$ and $1\PeV$, respectively
(25 bins), with $w=0.1$. The light spectrum is predicted as
$J_p+J_{\rm He}$, rather than fitting the derived helium points, which
inherit the anticorrelation of the subtraction. Both elemental
parameter sets are free, so charge ordering is tested rather than
imposed. The covariance combines independent statistical errors with an
illustrative within-spectrum systematic correlation
$\rho_{\rm sys}=0.5$, and neglects cross-covariances between the two
spectra, whose full form is unavailable.

Consistently with LHAASO findings, the joint broken-power-law fit gives
$E_{b,p}=(3.13\pm0.20)\PeV$ and
$E_{b,\rm He}=(7.2\pm0.8)\PeV$, with steepenings
$\Delta\gamma_p=0.85\pm0.06$ and $\Delta\gamma_{\rm He}=1.4\pm0.4$;
their ratio, $2.3\pm0.3$, is consistent with charge ordering and
insensitive to a shared energy scale. The two-cutoff hypothesis gives
$\chi^2/\nu=15.73/19$ with six free parameters, and the two-break
hypothesis $6.25/17$ with eight. For these non-nested hypotheses,
I calibrate $\Delta\chi^2_{\rm joint}=9.48$ with 5000 joint
pseudo-experiments generated under the fitted cutoff model, refitting
both families to each, yielding a preference for the broken-power-law
description with a $p$-value of approximately $0.004$.
The preference for a finite steepening is stable for
$\rho_{\rm sys}=0$--$0.9$ and similar for SIBYLL 2.3d.

Combining the elemental spectra with the all-particle intensity and
mean mass also constrains the nuclei beyond helium. At the all-particle
knee, $E=3.67\PeV$, the joint broken-power-law fit and the
EPOS-LHC all-particle reconstruction give $f_p\simeq0.35$ and
$f_{\rm He}\simeq0.29$: together they account for about $64\%$ of
the flux, leaving $f_X=1-f_p-f_{\rm He}\simeq0.36$.
Representing this remainder by one effective component, its mean
logarithmic mass follows from
\begin{equation}
 \langle\ln A\rangle_X=
 \frac{\langle\ln A\rangle-f_{\rm He}\ln4}
      {1-f_p-f_{\rm He}}
 \simeq2.36,
 \qquad A_{X,\rm eff}=\exp\!\big(\langle\ln A\rangle_X\big)
 \simeq10.6.
 \label{eq:knee-mass-balance}
\end{equation}
Here the all-particle intensity and $\langle\ln A\rangle\simeq1.25$
are interpolated to the knee energy~\cite{LHAASOall}; the elemental
fluxes come from the joint fit in Fig.~\ref{fig:kneeshape}.
The cutoff fit gives almost the same result, and the measured tables of
the three HIMs give $A_{X,\rm eff}\simeq10$--$12$.
A substantial unresolved component is therefore needed, with an
effective mass at or below that of carbon, markedly lighter than in
standard composition models
(e.g.\ $A_{X,\rm eff}\approx28$ in H3a~\cite{Gaisser2012}).
Since the $9\%$ systematic on $\langle\ln A\rangle$ is amplified by
$1/f_X$ to about $\pm0.3$ on $\langle\ln A\rangle_X$, this is a
consistency test between the KM2A composition and the hybrid elemental
fluxes rather than a bound: either medium--heavy nuclei are depleted at
the knee, or the two analyses differ in energy or flux scale.

In conclusion, the joint fit favors finite steepenings over exponential
cutoffs, with approximately charge-ordered break energies. The available
interval above each knee, especially for helium, is too short to
establish the asymptotic shape, and the preference depends on the HIM.
It therefore
constrains the shape to be reproduced by Galactic models, while leaving
open whether a transport transition or the combined cutoffs of a
source population produce it.

\section{Why a PeV scale?}
\label{sec:pev}

At high energies, $J\propto Q\,\tau_{\rm esc}$, so a knee can reflect
a source limit or a change in Galactic escape. The corresponding energy
scales can be estimated as follows:

\begin{itemize}
\item \textbf{Propagation.} Resonant scattering becomes less effective
when $r_{\rm L}\gtrsim\ell_c$, as little turbulent power is available
on larger scales. The diffusion coefficient can then grow faster with
energy, shortening $\tau_{\rm esc}\sim H^2/D$ in a halo of height $H$.
The condition $r_{\rm L}\sim\ell_c$ gives
\begin{equation}
 E_{\rm conf}(Z)\sim ZeB\ell_c
 \simeq3\,Z\PeV
 \left(\frac{B}{3\,\mu\mathrm{G}}\right)
 \left(\frac{\ell_c}{1\,\mathrm{pc}}\right).
 \label{eq:confinement}
\end{equation}
A knee at a few PeV thus requires $\ell_c\approx1\,\mathrm{pc}$, at the
low end of estimates for the coherence length of Galactic turbulence.
This transition is commonly associated with a change from Kolmogorov
diffusion, $D\propto E^{1/3}$, to small-angle scattering,
$D\propto E^2$~\cite{Kachelriess2019}. For an unbroken source spectrum
in a steady-state escape model, this implies a spectral steepening
$\Delta\gamma\simeq1.7$.

\item \textbf{Maximum source energy.} In the standard scenario in which
supernova remnants (SNRs) accelerate Galactic cosmic rays up to the knee,
the age-limited maximum energy is set by
$t_{\rm acc}(E_{\max})\simeq t_{\rm dyn}\sim R_{\rm sh}/u_{\rm sh}$,
typically near the onset of the Sedov--Taylor phase.
Assuming Bohm diffusion, $D_{\rm B}\simeq r_{\rm L}c/3$, the
acceleration time is $t_{\rm acc}\simeq\eta_{\rm acc}D_{\rm B}/u_{\rm sh}^2$,
with $\eta_{\rm acc}\sim20$ for a strong shock with equal upstream
and downstream diffusion coefficients. This gives~\cite{Blasi2013}
\begin{equation}
 E_{\max}(Z)\sim\frac{3}{\eta_{\rm acc}}\,
 ZeB_{\rm sh}R_{\rm sh}\frac{u_{\rm sh}}{c}.
 \label{eq:source-energy}
\end{equation}
The Galactic energy budget motivates $\xi_{\rm CR}\sim0.1$,
parametrized locally by $P_{\rm CR}=\xi_{\rm CR}\rho u_{\rm sh}^2$,
with upstream density $\rho$~\cite{Blasi2013}; if cosmic rays drive
magnetic amplification,
$B_{\rm sh}^2/(8\pi)=\xi_B\rho u_{\rm sh}^2$ with
$\xi_B\leq\xi_{\rm CR}$.
For a young, fast shock with $R_{\rm sh}\sim3\,\mathrm{pc}$,
$u_{\rm sh}\sim6000\,\mathrm{km\,s^{-1}}$, and
$\rho/m_p\sim0.1\,\mathrm{cm^{-3}}$,
\[
 E_{\max}^{\rm SNR}(Z)\sim3\,Z\PeV\,
 \frac{20}{\eta_{\rm acc}}
 \left(\frac{\xi_B}{0.1}\right)^{1/2}.
\]
In fact, realistic amplification saturates below this energy-budget
ceiling, making acceleration to PeV energies still challenging and highly
sensitive to the shock environment and evolutionary phase~\cite{Cristofari2021}.
\end{itemize}

Both estimates scale approximately with $Z$, reflecting their dependence
on rigidity $\rig\simeq E/(Ze)$. In a simple single-population picture,
the lower scale sets the knee. Distinguishing between an acceleration
limit and a change in Galactic escape is therefore a key step in
understanding the standard model of Galactic cosmic rays.

\section{Implications for Galactic cosmic-ray models}
\label{sec:models}

The detailed measurements discussed above reveal tensions with simple
descriptions based on a single power-law population and universal shapes
of the elemental rigidity spectra, with the knee attributed either to an exponential
source cutoff or to faster Galactic escape as scattering enters the
non-resonant regime. I briefly discuss recent works addressing these
issues.

\subsection{Knee ordering and spectral universality}

The approximate charge ordering of the proton and helium knees does not
require their full spectra to have the same rigidity dependence.
At relativistic energies, $E\simeq Ze\rig$, so Fig.~\ref{fig:ratio}
compares the two species at different total energies. A universal injection shape and common transport factor,
$J_i^{(\rig)}=a_i\,q(\rig)\,T(\rig)$ with fixed abundances $a_i$,
would cancel from this ratio; its evolution requires additional
structure in injection, composition, or source contributions.

Aharonian and Zhang~\cite{Aharonian2026} illustrate this distinction
with a phenomenological two-population model. The lower-energy proton
and helium components have different slopes and cutoff positions, while
the higher-energy helium component is constructed by rigidity scaling
of the proton component. Their superposition describes the main
spectral features and the changing proton-to-helium ratio from GeV to
PeV energies. This shows how a common rigidity dependence within one
population can coexist with different total elemental spectra.
The LHAASO measurements therefore test both the ordering of the knees
and the composition of the contributing populations~\cite{Dzhatdoev2026}.

\subsection{Acceleration limits and superposed Galactic contributions}

A source interpretation must reproduce the knee shape as well as its
energy scale. Superposed Galactic populations can generate the
preceding hardenings, while variations in maximum rigidity within each
population broaden its turnover. For sharp individual cutoffs and a
common injection index $\alpha$,
\begin{equation}
 J(\rig)\propto \rig^{-(\alpha+\delta)}\,S_w(>\rig),
 \label{eq:population}
\end{equation}
where $S_w$ is the luminosity-weighted fraction of contributing sources
able to reach $\rig$, and $D\propto\rig^\delta$.

\looseness=-1
In my recent population study~\cite{Evoli2026}, I showed that a
lognormal distribution of maximum energies produces a continuously
curving knee, whereas a power-law tail produces an approximately
constant post-knee steepening. A two-component fit to the proton
spectrum requires a logarithmic width
$\sigma_{\log_{10}E_{\max}}\simeq0.24$ for the PeV component.
Under the adopted SNR scalings, this is narrower than the dispersion
expected from explosion energies across the full remnant population,
favoring a restricted subset of PeV-capable SNRs or a different class of
sources. The preference for a
finite steepening over an exponential cutoff in
Section~\ref{sec:shape} motivates models with distributed maximum rigidities;
distinguishing a lognormal spread from a power-law tail requires spectra
extending further above the knees.

\subsection{Transport explanation}

A transport origin allows the sources to accelerate beyond the knee,
with the steepening imprinted by Galactic escape. The simple
Kolmogorov-to-small-angle transition in Section~\ref{sec:pev}
approaches $\Delta\gamma\simeq1.7$, larger than the $0.85$--$1.0$
measured for protons, unless the transition is broader than the
measured post-knee range; this motivates other transport mechanisms.

Our test-particle study~\cite{Drift2026} examines the competition
between diffusion and magnetic drifts through both escape times and
grammage. A purely azimuthal regular field with Kolmogorov turbulence
produces a knee, but perpendicular escape is too slow and the grammage
is excessive. Adding a vertical field component permits faster parallel
escape, which suppresses the drift-induced break. If the parallel
diffusion coefficient instead becomes nearly energy independent above
TeV energies, drifts can dominate at PeV energies and produce a knee
while matching the TeV grammage normalization. Under these assumptions,
the transition to drift escape, $\tau_{\rm esc}\propto\rig^{-1}$, can
produce $\Delta\gamma\simeq1$, close to the measured proton steepening.

\section{Conclusions and the next discriminating measurements}
\label{sec:conclusions}

\looseness=-1
LHAASO turns the first knee into a joint test of spectrum and composition.
The all-particle knee is driven mainly by protons, and the proton and
helium knees, at $\simeq3.1$ and $\simeq7\PeV$, are compatible with a
common rigidity, while the proton-to-helium ratio at equal rigidity,
rising toward the knee, challenges a universal shape with fixed
abundances. For two of three HIMs the knees are better described by a
finite steepening than by exponential cutoffs, motivating models with
distributed maximum rigidities, and the nuclei beyond helium are lighter than in
standard composition models. Whether the Galactic component extends to
an iron knee or gives way to an extragalactic one soon after the light
knees~\cite{Dzhatdoev2026} depends on these heavier nuclei.
Measurements above the helium knee, elemental spectra beyond helium,
documented covariances, and overlap with direct experiments are
therefore priorities, while anisotropy and diffuse emission will test
whether the local knee represents the
Galaxy~\cite{Diffuse2025,Prevotat2025}.

\clearpage
\setlength{\bibsep}{0pt}

\end{document}